# INTERROOM RADIATIVE COUPLINGS THROUGH WINDOWS AND LARGE OPENINGS IN BUILDINGS : PROPOSAL OF A SIMPLIFIED MODEL

H. Boyer[1], M. Bojic[2], H. Ennamiri[1], D. Calogine[1], S. Guichard[1]

[1] University of La Réunion, PIMENT Lab., France

[2] University of Kragujevac, Faculty of Mechanical Engineering, Serbia

Corresponding author : harry.boyer@univ-reunion.fr

## ABSTRACT

A simplified model of indoor short wave radiation couplings adapted to multi-zone simulations is proposed, thanks to a simplifying hypothesis and to the introduction of an indoor short wave exchange matrix. The specific properties of this matrix appear useful to quantify the thermal radiation exchanges between the zones separated by windows or large openings. Integrated in CODYRUN software, this module is detailed and compared to experimental measurements carried out on a real scale tropical building.

## NOMENCLATURE

Variables

| | |
|---|---|
| $A$ | area (m²) |
| $D$ | direct solar radiation (W.m$^{-2}$) |
| $d$ | diffuse solar radiation (W.m$^{-2}$) |
| $\phi$ | flux (W) |
| $F_{ij}$ | view factor |
| $\rho$ | short wave reflectivity |
| $\alpha$ | short wave absorptivity |
| $\tau$ | window short wave transmitivity |
| $r$ | leak ratio of a zone |
| $a_{ij}$ | term of zone coupling matrix |

Subscripts :

| | |
|---|---|
| $D$ | direct |
| $d$ | diffuse |
| $a$ | absorbed |
| $r$ | received |
| $T$ | total |
| $W$ | window |
| $O$ | opaque walls |

# 1. Introduction

Multi-zone software accounts for more and more complex phenomena as airflow transfers, lighting, pollutants, and and so on. Meanwhile, information concerning specific parts of the global model is not always available in literature.

The direct and diffuse solar radiation entering a building through windows and openings of internal impact surfaces, and after reflections, is absorbed by these surfaces or escape to the outside through the windows or openings. This last part can be important in the case of highly glazed buildings and this was the subject of previous studies - emphasizing on the simulation codes, which do not always reflect this.

The concept of solar absorptance - defined as the ratio between the energy absorbed and the incident - was initially applied to the study of solar captation [8] and was numerically studied by [6] in the case of a single zone, the object being to identify simplified models of computation. Detailed models based on complete geometrical description of enclosures with respect to dimensions are available for mono-zone cases [7, 15, 18, 17].

If models are available for previous cases, not many details are given in literature on short wave repartition and zone couplings. A detailed geometrical description of the building allows a detailed calculation of indoor thermal radiation couplings, but it leads to a large amount of calculation that can be avoided.

This paper deals with indoor short wave radiation exchanges between zones through windows and large openings. The importance of these transfers is obvious as regarding lighting. But another aspect is their impact on indoor dry-bulb temperatures. It is well-known that airflow transfers, which play a very important role in terms of energy needs or thermal comfort, are very sensitive to the gradient of air temperature and it is therefore crucial to properly model the thermal radiation exchange.

This paper reports a simplified approach to the calculation of indoor thermal radiation couplings, which saves the calculation time.

## 2. Indoor short wave repartition modeling

### 2.1 Initial model in CODYRUN

When the short wave exchanges are not taken into account, direct $D$ and diffuse $d^0$ solar radiations represent the solar inputs (in W) within a zone (thanks to the external windows and lighting). In the calculation process, superficial indoor nodes are affected with an algorithm repartition (direct beam has often had an incident on slabs and diffuse radiation is identically received by all the indoor surfaces). It is to be noticed that multiple reflection is not often considered in common software as TRNSYS [11].

The initial model in CODYRUN comes from the book of algorithms : CODYBA [4]. This monozone model calculates the solar radiation absorbed by the indoor walls from incident direct and diffuse radiation in a specific zone. The indoor envelope of a zone is supposed to be composed of four entities: slabs, vertical walls, internal walls and outdoor windows. They are subscripted from 1 to 4 respectively in the next equations. As an example, the wall entity aggregates the real walls of the zone and is affected by the sum of their surfaces and average reflectance (weighted by the real surfaces of the walls). The advantage of this procedure is to reduce to four the size of the system to solve.

In the zone, a first assumption concerns the repartition of the incident radiation. The direct part is supposed to be an incident on the slab(s) whereas the diffuse radiation is uniformly shared by the entities, in regard to their surfaces. The view factors are calculated according to spherical approximation. $A_T$ being the total indoor surface, incident fluxes $\phi_i^0$ on each entity can be calculated from direct ($\phi_D$) and diffuse ($\phi_d$) incident radiation through the following relations:

$$\begin{Bmatrix} \phi_1^0 \\ \phi_2^0 \\ \phi_3^0 \\ \phi_4^0 \end{Bmatrix} = [M] \begin{Bmatrix} D \\ d^0 \end{Bmatrix}$$

$$\text{with} \quad [M] = \frac{1}{A_T} \begin{bmatrix} A_T & A_1 \\ 0 & A_2 \\ 0 & A_3 \\ 0 & A_4 \end{bmatrix}$$

In order to take into account multiple reflections, from each previously defined entity, the total solar radiation received, $\phi_r$ is introduced. Thus,

$$\phi_{r,i} = \sum_{j=1}^{4} F_{ij} \rho_j \left( \phi_j^0 + \phi_{r,j} \right), \text{ with view factor } F_{ij} = \frac{A_j}{A_T}$$

The absorbed radiation is then

$$\phi_{a,i} = \alpha_i \left( \phi_i^0 + \phi_{r,i} \right) = \alpha_i \phi_i^0 + \alpha_i \sum_{j=1}^{4} F_{ij} \rho_j \left( \phi_j^0 + \phi_{r,j} \right)$$

that is $\sum_{j=1}^{4} F_{ij} \frac{\rho_j}{\alpha_j} \phi_{a,j} - \frac{1}{\alpha_i} \phi_{a,i} = -\phi_i^0$

In a matrix form, $[M_a] \phi_a = -[\phi_i^0]$

At last, the absorbed flux $\phi_a$ is

$$\phi_a = -[M_a]^{-1}[M] \begin{Bmatrix} D \\ d^0 \end{Bmatrix}$$

with

$$M_a = \begin{bmatrix} \frac{\rho_1}{\alpha_1} F_{11} - \frac{1}{\alpha_1} & \frac{\rho_2}{\alpha_2} F_{12} & \frac{\rho_3}{\alpha_3} F_{31} & \frac{\rho_4}{\alpha_4} F_{41} \\ \frac{\rho_1}{\alpha_1} F_{12} & \frac{\rho_2}{\alpha_2} F_{22} - \frac{1}{\alpha_{22}} & \frac{\rho_3}{\alpha_3} F_{32} & \frac{\rho_4}{\alpha_4} F_{42} \\ \frac{\rho_1}{\alpha_1} F_{13} & \frac{\rho_2}{\alpha_2} F_{23} & \frac{\rho_3}{\alpha_3} F_{33} - \frac{1}{\alpha_3} & \frac{\rho_4}{\alpha_4} F_{43} \\ \frac{\rho_1}{\alpha_1} F_{14} & \frac{\rho_2}{\alpha_2} F_{24} & \frac{\rho_3}{\alpha_3} F_{34} & \frac{\rho_4}{\alpha_4} F_{44} - \frac{1}{\alpha_4} \end{bmatrix}$$

For this model, the zone is represented by a spherical enclosure - each of the entities being part of the surface. On the next figure, the entity aggregating slabs, which are supposed to receive the direct beam, have been grayed.

**Fig. 1 : Spherical view of a zone.**

## 2.2 The new proposal model

The first objective is to obtain the amount of outgoing radiation through a window, by taking into account the indoor reflection.

**Fig. 2 : Basic problem**

A simplified equation of the transmitted flux $\phi$ through the internal window (or a large opening) can be obtained. Let us denote $A_w$ the window surface, $A_o$ the opaque surface of the walls and $A_T$ $(= A_O + A_W)$ the total indoor surface. Assuming that the diffuse radiation is isotropic, the incident part on the window is $d^0 \frac{A_w}{A_T}$.

Considering the average short wave reflectance of walls, they reflect $\rho_o d^0 \frac{A_o}{A_T}$ for the first reflection and the window intercepts $\rho_w d^0 \frac{A_o}{A_T} \frac{A_w}{A_T}$.

Continuing the calculation process, the window receives

$$d^0 \frac{A_w}{A_T}(1+\rho_o \frac{A_o}{A_T}+(\rho_o \frac{A_o}{A_T})^2+...) = d^0 \frac{A_w}{A_T}\frac{1}{1-\rho_o \frac{A_0}{A_T}}$$

That, $\phi = \tau_d \, d^0 \frac{A_w}{A_T}\frac{1}{1-\rho_o \frac{A_0}{A_T}}$ where $\tau_d$ is the window diffuse transmittance (the value of 1 will be used in case of a large opening). If we assume that the direct beam $D$ is received by the floor and reflected in a diffuse way, the following expression is obtained:

$$\phi = \tau_d \left[ d^0 + \rho_{floor} \, D \right] \frac{A_w}{A_T} \frac{1}{1-\rho_o \frac{A_T - A_w}{A_T}}$$

If $\rho_o$ equals zero (black walls), the obtained value only depends on the view factor. One can notice that the reflectance of the walls tends to increase $\phi$.

## 3. Matrix approach of couplings :

The multi-zone building simulation codes need to integrate a detailed short wave coupling model including outdoor exchanges. Because of *the non-geometrical building description*, the hypothesis of no indoor direct beam couplings can be made - what is common in a real building.

### 3.1 Indoor short wave couplings :

A matrix is introduced, traducing Zone Diffuse Couplings (ZDC). Each term $a_{ij}$ of this matrix defines the rate of transfer from zone $i$ to $j$. $k$ indexes a window between the previous zones.

$$a_{ij} = \sum_k \tau_{dk} \frac{A_{wk}}{A_{Ti}} \frac{1}{1-\rho_o(\frac{A_{Ti}-A_{wk}}{A_{Ti}})}$$

This term depends on the surfaces, on the colour of the indoor walls (through $\rho_o$) and on the diffuse window transmittance $\tau_d$. An approximate value is $\tau_{dk} \frac{A_{wk}}{A_{Ti}}$, corresponding to all indoor surfaces being black ($\rho_o$ = 0). Some interesting remarks can be made, considering two zones indexed by *i* and *j* :

- $a_{ij} \in [0,1[$ (typical value is 1/50)

- for 1W in zone *i*, $a_{ij}$ W will be transferred to *j*.

- diagonal terms are equal to zero.

- $a_{i0}$, is linked to the outdoor windows and openings, and expresses short wave "leaks"

- $r_i = \sum_j a_{ij}$ gives the leak ratio of zone *i*.

If $r_i \approx 1$, the radiation entering is nearly transferred to other zones (including outdoor). An example is the air cavity in double glass windows or highly glazed buildings. For the reverse case, the radiation is mainly trapped in the room (solar captation enclosure in solar heaters). This can easily be linked to what authors call *room absorbance* [6]. Continuing the same representation as for the initial single zone model, the following scheme is obtained :

**Fig. 3 : Spherical view of the couplings.**

**3.2 Algebraic formulation of the couplings in multizone cases :**

A matrix writing of the short-wave radiation exchanges between the zones is then obtained. The aim is to modify the incident diffuse radiation into each zone to obtain a net diffuse radiation. Then, as previously, the surface repartition is done as in the initial single zone model (2.1).

Let us consider three zones indexed by 1, 2 and 3, with no external radiation exchanges (i.e. no external windows or large openings). For zone $i$, $d_i^0$ is the incident diffuse radiation in the zone $i$ (in that case, diffuse radiation can be due to internal lighting) and $d_i$ the net diffuse radiation resulting from the internal couplings. In the thermal equations of zone $i$ (more precisely in thermal radiation balance on indoor surfaces), this net variable is the one to be used and has to be determined. The radiation balance of the first zone, taking into account the indoor coupling terms $a_{ij}$, leads to

$$d_1^0 + \rho_1 D_1 = d_1 + (a_{12} + a_{13})d_1 - a_{21} d_2 - a_{21} d_3$$

The first term is equal to the radiation available in the room before inter-zone exchanges. The second expresses the same amount, but involves the diffuse radiation and integrates a net exchange.

That can be rewritten under the following form

$$(1 + a_{12} + a_{13}) d_1 - a_{21} d_2 - a_{31} d_3 = d_1^0 + \rho_1 D_1$$

If we now consider the 3 zones:

$$\begin{bmatrix} 1+a_{12}+a_{13} & -a_{21} & -a_{31} \\ -a_{12} & 1+a_{21}+a_{23} & -a_{32} \\ -a_{13} & -a_{23} & 1+a_{31}+a_{32} \end{bmatrix} \begin{pmatrix} d_1 \\ d_2 \\ d_3 \end{pmatrix} = \begin{pmatrix} d_1^0 + \rho_1 D_1 \\ d_2^0 + \rho_2 D_2 \\ d_3^0 + \rho_3 D_3 \end{pmatrix} \quad (1)$$

Some coherent tests has succeeded, when the direct radiation is not incident in the rooms :

- The single zone case leads to $d_1 = d_1^0$

- The sum of the three equations gives $d_1 + d_2 + d_3 = d_1^0 + d_2^0 + d_3^0$, what points out the short wave indoor radiation conservation (due to the non-existence of external coupling terms)

- Without exchanges (i.e. coupling terms), $a_{ij} = 0$, and M = $I_d$ (Identity matrix) and $d_i = d_i^0$

Indexing the outside by 0, adding external (i.e. terms $a_{i0}$), terms to Eq. 1 we have

$$\begin{bmatrix} 1+a_{10}+a_{12}+a_{13} & -a_{21} & -a_{31} \\ -a_{12} & 1+a_{20}+a_{21}+a_{23} & -a_{32} \\ -a_{13} & -a_{23} & 1+a_{30}+a_{31}+a_{32} \end{bmatrix} \begin{pmatrix} d_1 \\ d_2 \\ d_3 \end{pmatrix} = \begin{pmatrix} d_1^0 + \rho_1 D_1 \\ d_2^0 + \rho_2 D_2 \\ d_3^0 + \rho_3 D_3 \end{pmatrix}$$

Leading to $M\,d = N$, with  (2)

$$M[i,i] = 1 + \sum_{j=0}^{j=n,\,j!=i} a_{ij}$$

$$M[i,j] = -a_{ji},\ for\ i \neq j$$

$$N[i] = d_i^0 + \rho_i\,D_i$$

As a conclusion, the solution of equation (2), using the ZDC matrix, allows/permits the determination of the net diffuse radiation in each zone resulting from the couplings.

Coherent tests were successfully conducted, using simple geometry and cases (black walls, white walls), and confirm the interest of this approach.

### 3.3 Integration in CODYRUN software :

The previous model was integrated in CODYRUN. This application is mainly used in the passive architecture design [1, 2, 3, 10]. Large indoor openings have been previously integrated, but without taking into account radiation exchanges. After making modifications in the source code, some new outputs were added, as the radiation fluxes has been transferred from each/one zone to others, allowing a detailed analysis of energy flow paths.

### 4. Experimental validation :

### 4.1 The test case

A typical dwelling of collective housing in Reunion island has been monitored. This assessment is part of a technical evaluation of building standards [10] for French overseas territories. The dwelling, represented in Fig. 4, includes three bedrooms and a living room. Large bay windows are present on the West side of the dwelling (see also Fig. 5).

The measurement program lasted some time in January 1998. As for the meteorological data, the equipment was set-up on the terrace roof of the dwelling, giving half-hourly data for outdoor dry-bulb temperature, relative humidity, global and diffuse horizontal radiation as well as wind speed and direction.

The ambient temperatures (dry-bulb and resultant) and relative humidity were measured in two places and at different heights in each room of the dwelling. Furthermore, some indoor walls and roof surface temperatures were included in the experimental results. Some elements linked to the measurements on this dwelling are available in refs [1,13]. For this specific study, the vertically transmitted radiation was also measured. Other papers related to CODYRUN software validation are covered by [5,9,14,16].

Fig.4 : Typical dwelling

In terms of the dry-bulb temperature, a good agreement of simulation with measurement results is obtained, as displayed in the following graph showing the residual zone of the *Living Room* (difference between the dry temperature measures calculated, in K) for a five-day period.

Fig. 5 : *Living Room* **dry-bulb residual temperature**

**4.2 Application of the coupling matrix zone to the case :**

Our building description affects the numbers 1 for *Room1* zone, 2 for *Room2*, 3 for *Room3* and 4 for the *Living room*.

During the simulations, an intermediate output of the code is the zone coupling matrix associated to this building (rounded to two digits).

$$[a_{ij}] = \frac{1}{100} \begin{pmatrix} 0 & 0 & 0 & 0 & 0 \\ 7 & 0 & 0 & 0 & 0 \\ 15 & 0 & 0 & 21 & 0 \\ 10 & 0 & 0 & 0 & 0 \\ 10 & 0 & 8 & 0 & 0 \end{pmatrix}$$

For example, the term $a_{24}$ (equal to 0.21) expresses the exchanges between the *Room2* zone and the *Living room*, through the open sliding door. It can be noticed that 21% of incident radiation in *Room2* is transferred into the *Living room*. This important value is linked to the large surface of the opening (3.45 m², see Fig. 4. and 9).

By adding the terms of line 2 in previous matrix, 36% of the incident radiation in *Room2* is transmitted to the outside and to zone 4 through the terms $a_{24}$ and $a_{20}$.

### 4.3 Comparison of the incoming and outgoing solar radiation

A first necessary step was to check the incoming and outgoing solar radiation into each zone before treating the indoor radiation exchanges.

**Fig.6 : Experimented dwelling on its last floor (West side)**

For this sequence, two pyranometers were installed on the west glass bay window(s) (of the living room) and another one can be seen on the previous picture. Thus, we compare the global exterior solar radiation with the interior one, both for the actual measurements and for the model predictions. The manufacturer indicates the error to be of 5%, for the used pyranometer (Kipp and Zonen, model CM 6B). CODYRUN had already been tested with respect to the effect of shading devices on the direct radiation with use of BESTEST procedure [16]. In ref. [13], when comparing with measurements, it appears that the model was overestimating from 30 to 40% of the interior radiation contributions in the living room (behind the bay window). Furthermore, in the same paper, the sensitivity analysis has shown that the transmittance of the glass was not an important factor. The near shading device that can be seen on fig. 7, is an egg-crate composed of overhang, left and right fins, around the west bay window was clearly responsible for the overestimation of the solar gains.

**Fig. 7 : Near shading device detail**

A first step was then to improve the model by introducing the possibility of reducing the diffuse solar radiation with near shading devices [13], allowing to obtain the following curves:

**Fig. 8 : Comparison of the incoming global solar radiation**

The relative average daily error in the incoming radiation (before bay window) is about 4% over the whole period (see Fig. 8), as regards the same order of magnitude of the measurement error. Even if CODYRUN succeeded the IEA BESTEST procedure [16], discrepancies can appear between the curves and they are linked to additional uncertainties on measurements (for example some difficulties can appear to set the pyranometer in the same plane of the bay window) or the modelling complex shadow mask seen on fig. 7. Meanwhile, towards our objectives, the global incident radiation into the zone can be considered as correctly evaluated.

A specific comment can be made concerning the incoming radiation peak measurement, on day 2 for example. Using a single pyranometer, due to the overhang, this sensor switches from shaded to direct exposed and measurement curve exhibits a strong peak at this moment. The model - which considers/(takes into account) at each moment the fraction of the bay window shaded to compute the solar mask effect on the solar input - does not show such a peak.

Figure 9 shows the outcoming solar radiation from escaping fron Living room through oudoor bay window. Here, the same remarks can be made. This shows that more uncertainties, which are linked to the model of the indoor repartition radiation, can be encountered.

**Fig. 9 : Comparison of the outcoming global solar radiation measurement with CODYRUN output**

In the afternoon, during sunny days (days 2 and 3), the values of the measured peaks (in the order of 20 $W.m^{-2}$) are not reproduced by the model. They are related to the penetration of direct radiation into the living room (west-side exposed) in the afternoon. More precisely, for these two days, in the afternoon, if the peak is reproduced in the model response, an adjustment of the floor reflectivity is clearly needed.

**4.4. Comparison of short wave couplings**

A specific 3 days sequence was instrumented with the two pyranometers located at the section of the open sliding door (1.6 x 2.1 m) between the Living room and Room 2 (see

Fig. 10). In this figure, sensors for the mean radiant temperatures and dry-bulb can also be seen, as the data.

**Fig. 10 : Monitoring of indoor short - wave radiation exchange**

Comparisons between the short-wave radiation exchange (between the two zones) and measurements can be made. Having only one measurement point in the sliding door, it is necessary to make a hypothesis that the value of the short-wave radiation exchange is constant in the entire surface.

**Fig. 11 : Comparison of Measurement/Model for the radiation heat flux transmitted from the *Living Room* to *Room2***

The measured value seems quite constant, except at the end of days when the direct solar radiation enters the living room, due to an external bay exposal (West). When analysing the heat radiation exchanges from the *Living room* to *Room2*, at the sunset, the incident direct radiation in the living room is detected by the sensor.

Globally, an agreement with the model is not as good as previously. From the model point of view, it might point out failings of the simplifying assumptions in our model: no direct couplings, diffuse indoor reflections, approximations of view factor and aggregation of components in four entities (cf 2.1). From the experimental point of view, this comparison also suffers from large uncertainties in the indoor surface reflectance. At the end, the use of a single measurement point in this large enclosure is also problematic.

However, the presented curves show that this approach takes into account these couplings, at least, to estimate their magnitudes at each time step.

## 5. Conclusion

The detailed geometrical description of the building allows a detailed calculation of the indoor thermal radiation couplings, but leads to a large amount of calculation that can be avoided. The simplified proposed method shows its applicability to evaluate the outdoor/indoor exchanges, but shows some limits when strong direct thermal radiation couplings exist.

It is important to note that the matrix of inter-zone transfer can be calculated without any simulation tool and allows a better understanding of the thermal radiation exchange and HVAC sizing.

A step-by-step approach has been used to improve and contribute to the validation of the thermal calculations by using the simulation program. Further works should be done to improve the modelling aspects and measurements. The first task could be to avoid aggregation of components in order to obtain more realistic repartition of radiation, and a second one to conduct more detailed instrumentation (i.e. multiple point radiation measurements in the enclosures). Meanwhile, this approach is better than the common non-coupling assumption and leads to improve knowledge about energy flow paths in buildings and systems, as for example in partially glazed Trombe wall [12], the case in which direct radiation couplings and mass transfers occur simultaneously.

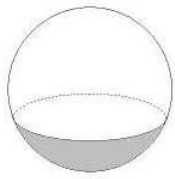

**Fig. 1 : Spherical view of a zone.**

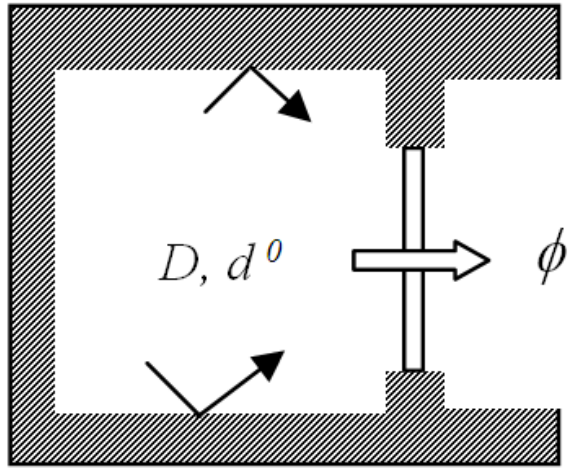

**Fig. 2 : Basic problem**

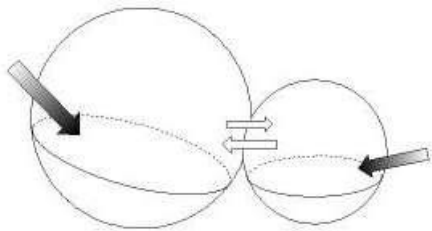

**Fig. 3 : Spherical view of the couplings.**

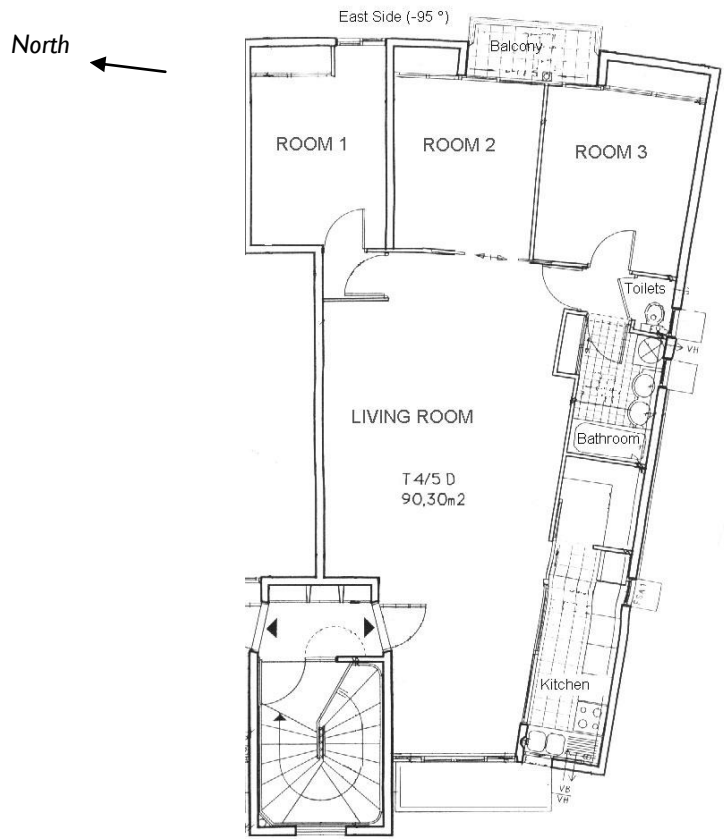

**Fig.4 : Typical dwelling**

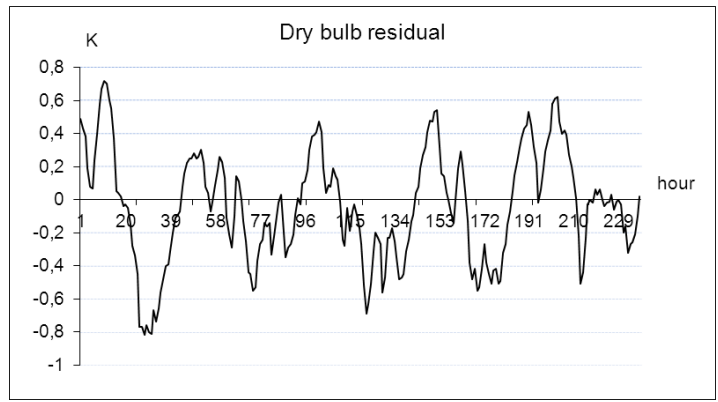

**Fig. 5 :** *Living Room* **dry-bulb residual temperature**

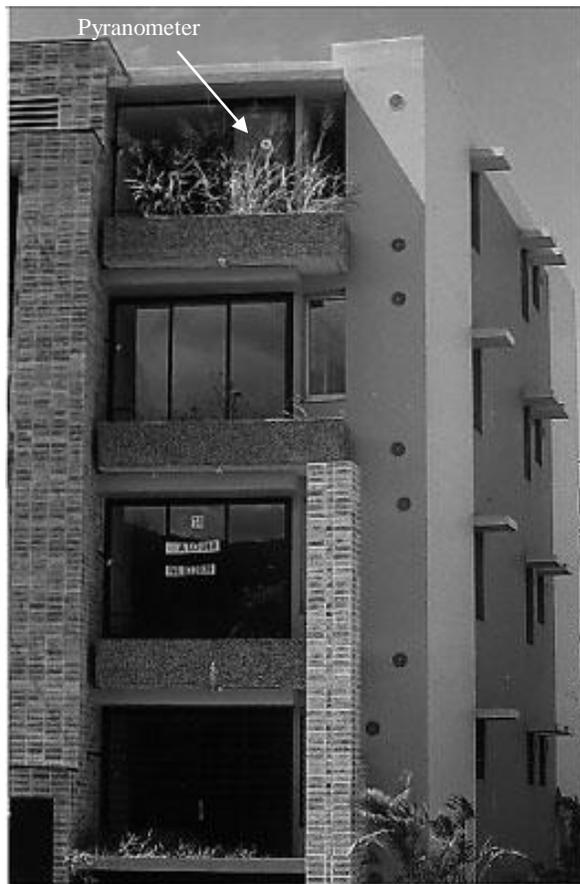

**Fig.6 : Experimented dwelling on its last floor (West side)**

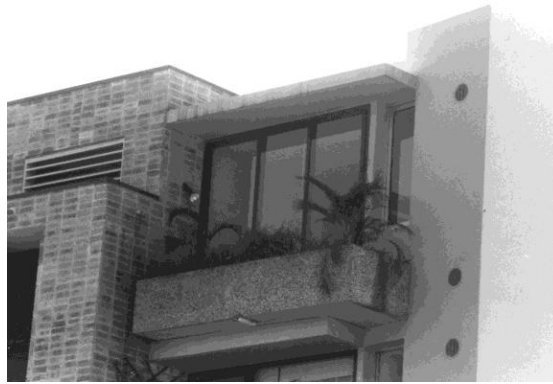

**Fig. 7 : Near shading device detail**

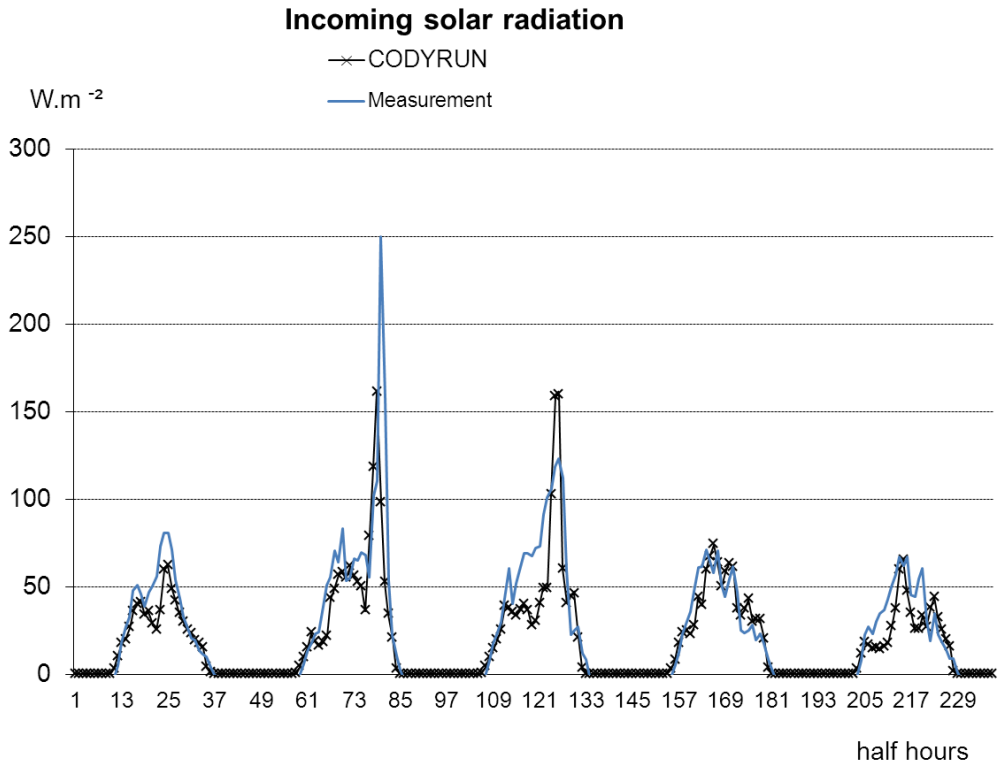

**Fig. 8 : Comparison of the incoming global solar radiation**

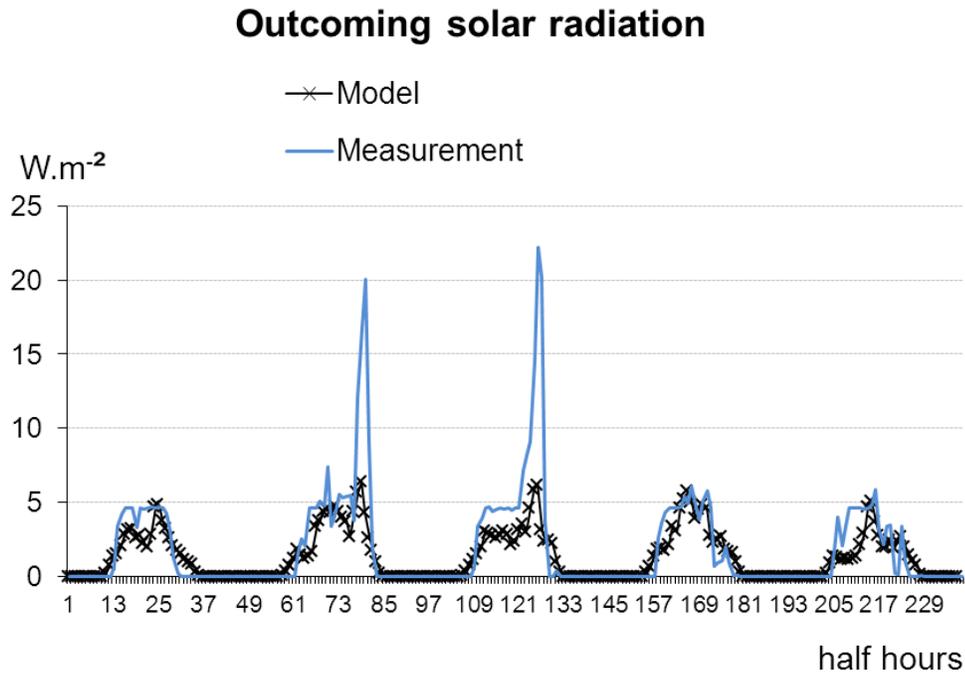

**Fig. 9 : Comparison of the outcoming global solar radiation measurement with CODYRUN output**

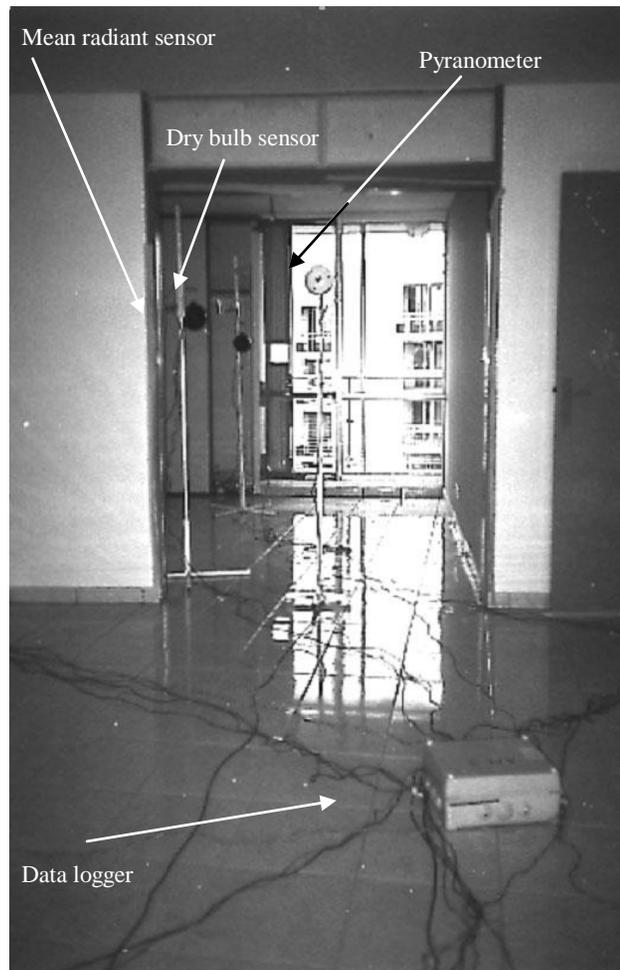

**Fig. 10 : Monitoring of indoor short - wave radiation exchange**

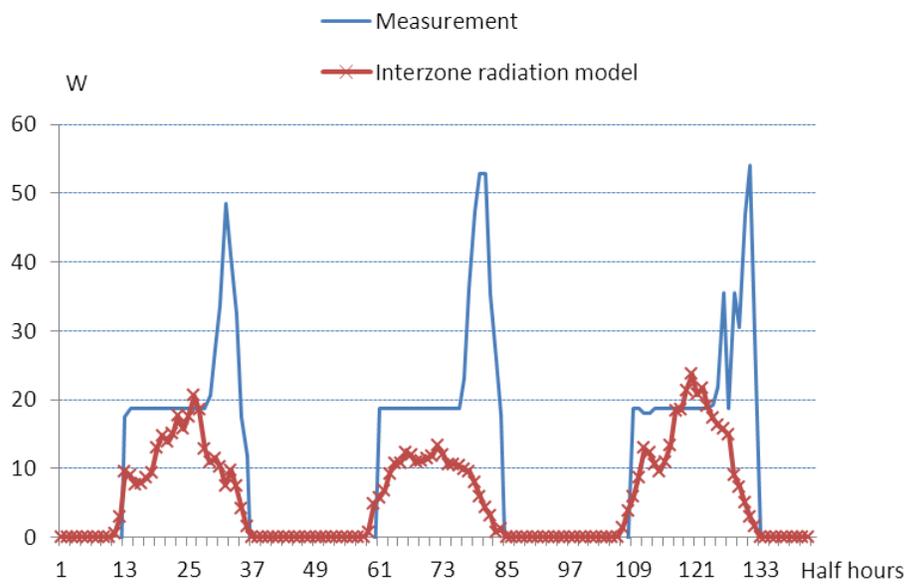

**Fig. 11 : Comparison of Measurement/Model for the radiation heat flux transmitted from the *Living Room* to *Room2***